# GeoLocator: a location-integrated large multimodal model for inferring geo-privacy

Yifan Yang[1], Siqin Wang[1], Daoyang Li[1], Yixian Zhang[2], Shuju Sun[1], Junzhou He[3]

[1] Spatial Sciences Institute, University of Southern California, Los Angeles, US.

Y.Y.: yyang295@usc.edu; S.W.: siqinwan@usc.edu; D.L.: daoyangl@usc.edu; S.S.: shujusun@usc.edu

[2] Evolutionary Assets, Shenzhen, China. Y.Z.: zhangyx@jhlfund.com

[3] Viterbi school of engineering, University of Southern California, Los Angeles, US. junzhouh@usc.edu

**Corresponding author:**

Yifan Yang, Spatial Sciences Institute, University of Southern California, Los Angeles, US. yyang295@usc.edu

Siqin Wang, Spatial Sciences Institute, University of Southern California, Los Angeles, US. siqinwang@usc.edu

## Abstract

Geographic privacy or geo-privacy refers to the keeping private of one's geographic location, especially the restriction of geographical data maintained by personal electronic devices. Geo-privacy is a crucial aspect of personal security; however, it often goes unnoticed in daily activities. With the surge in the use of Large Multimodal Models (LMMs), such as GPT-4, for Open Source Intelligence (OSINT), the potential risks associated with geo-privacy breaches have intensified. This study develops a location-integrated GPT-4 based model named *GeoLocator* and designs four-dimensional experiments to demonstrate its capability in inferring the locational information of input imageries and/or social media contents. Our experiments reveal that *GeoLocator* generates specific geographic details with high accuracy and consequently embeds the risk of the model users exposing geospatial information to the public unintentionally, highlighting the thread of online data sharing, information gathering technologies and LLMs on geo-privacy. We conclude with the broader implications of *GeoLocator* and our findings for individuals and the community at large, by emphasizing the urgency for enhanced awareness and protective measures against geo-privacy leakage in the era of advanced AI and widespread social media usage.

# 1. Introduction

In today's digital era, the silent leakage of personal information is a growing concern, with geographic privacy being a pivotal area of focus. Geographic privacy pertains to the protection and confidentiality of geographic information linked to individuals. It primarily encompasses safeguarding data that discloses an individual's geographic location, such as real-time whereabouts, historical movement patterns, or any location-specific information that can be traced back to them. The significance of geographic privacy is paramount. However, maintaining this privacy poses a significant challenge in the age of ubiquitous smartphones and social media platforms. While services like navigation, travel ticketing sites, and social media offer convenience, they simultaneously risk compromising our geographic privacy through potential surveillance, unauthorized data mining, and third-party misuse. This concern is further amplified by the existing legal framework's inability to keep pace with the rapidly evolving technologies that threaten geographic privacy.

We have discovered an easily overlooked way to give away our geographic privacy, and our everyday photos often contain a lot of geographic privacy information. For example, we post a photo on social media showing ourselves visiting a particular ballpark. We can infer your geographic location from this photo. Specifically, which ballpark is in your photo and what is the specific address information for that ballpark. One photo is enough to give away your geographic privacy. With the rapid development of large multimodal models such as GPT-4, which is capable of extracting, interpreting, and inferring geographic information from your published images, the geographic information is enough to expose your privacy. The ability that GPT-4 has to infer photos pose a significant threat to geographic privacy. These models have the potential to reveal precise location details directly from geotagged images or indirectly through contextual analysis. The potential harms and implications are far-reaching and multifaceted, including identity theft, personal security breaches, and the risk of serious intrusions into an individual's private life.

Recognizing the potential threat posed by GPT-4 to geo-privacy, we developed *GeoLocator*, a tool integrating GPT-4 with geolocation function, and demonstrate its capability in inferring the locational information of input imageries and/or social media contents. To evaluate and compare the capability of regular search engines, GPT-4, and *GeoLocator* in perpetrating privacy attacks, we design a series of experiments in four perspectives based on the input of various datasets, including Google Maps images, daytime/nighttime images, and social media posts. Our experiments reveal that *GeoLocator* generates specific geographic details with high accuracy and consequently embeds the risk of the model users exposing geospatial information to the public unintentionally, highlighting the thread of online data sharing, information gathering technologies and LLMs on geo-privacy. We conclude with the broader implications of *GeoLocator* and our findings for individuals and the community at large, by emphasizing the urgency for enhanced awareness and protective measures against geo-privacy leakage in the era of advanced AI and widespread social media usage.

## 2. Related Work

We commence with providing a comprehensive overview of the capabilities of Large Multimodal Models (LMMs) in attacking geo-privacy, and the key milestones and innovative techniques that have shaped the evolution of LMMs.

### 2.1 LMMs introduction

The transformative emergence of the Transformer architecture (Vaswani et al., 2017) set a new precedent in the field, laying a robust foundation for contemporary large language models. This breakthrough was followed by the development of pivotal models in Natural Language Processing, notably GPT (Radford et al., 2018) and Bidirectional Encoder Representations from Transformers (Devlin et al., 2018). More recently, with the development of computing power and advanced training techniques such as instruction tuning and reinforcement learning from human feedback (Ouyang et al., 2022; Wang et al., 2022), LLMs such as ChatGPT (OpenAi, 2023) can achieve superior results in various downstream applications without the need for task-specific tuning. For example, LLMs excel in abstract summarization, producing meaningful overviews of text passages. This capability can be particularly beneficial in fields with vast amounts of text, like legal practice, academic research, and medicine, aiding in the efficient navigation of dense information repositories (Holmes et al., 2023; Yuan et al., 2023). Furthermore, LLMs have the ability to understand context and user intent has led to applications in customer service, personal assistance, and interactive educational tools (Hou et al., 2023).

Concurrently, there is a notable trend towards integrating LLMs with vision-based models, heralding a new era of Large Multimodal Models. This integration expands the range of tasks they can perform and aligns more closely with the multimodal nature of human cognition. LMMs differ from LLMs by processing and interpreting both textual and other types of data such as images. This advancement led to groundbreaking advancements in visual understanding and reasoning. For instance, the proprietary GPT-4 model (OpenAi, 2023), renowned for its illustrative abilities, and open-source models like Large Language and Vision Assistant (Liu et al., 2023), have demonstrated exceptional skill in blending textual and visual information. These models have shown proficiency in tasks ranging from generating website code from visual prompts (Zhu et al., 2023) to recognizing complex details in image-rich contexts (Zhang et al., 2023). Their success illustrates not only the versatility of LMMs in handling multimodal data but also their potential in transforming tasks that require an intricate understanding of both visual and textual elements. In the next section, we continue to explore existing work integrating large multimodal models with a variety of tasks and examine the application of artificial intelligence in geography.

### 2.2 LMMs applications

LMMs have a very wide range of application capabilities. LMMs are the most cutting-edge technology that has been widely employed in diverse domains. In the medical field, Hou et al. found that the current multimodal models, GPT-4 and Bard, could handle visual assignments. For example, GPT-4 successfully solved 96.7% of the visual problems, facing minimal difficulty with only one Parsons problem (Hou et al., 2023). Yuan et al. found that LMMs can be applied to enhance various aspects of healthcare. Particularly, it highlights the crucial role of LMMs, investigating their ability to process diverse data types like medical imaging and Electronic Health Records to augment diagnostic accuracy (Yuan et al., 2023). Fabian et al. proposed a novel zero-shot species classification framework that leverages multimodal foundation models. This framework involves instruction tuning vision-language models to generate detailed visual descriptions of camera trap images, using terminology similar to that of experts (Fabian et al., 2023). Picard et al. evaluated GPT-4V, a vision language model, in engineering design tasks, demonstrating its capabilities and limitations. Their study provides foundational insights for the application of vision language models in engineering (Picard et al., 2023). Warner et al. explored the shift in medical AI systems towards deep learning models, focusing on LMM's impact on medical image analysis and clinical decision support systems (Oh et al., 2023). Oh et al. introduced an LMM for radiation therapy, integrating clinical text with images, demonstrating enhanced performance in breast cancer treatment, a first in such clinical text integration for oncology (Yang et al., 2023). Microsoft delved into the capabilities of GPT-4Vision, highlighting its proficiency in video understanding, visual reasoning, and other areas. They underscored the substantial potential applications of this technology in various sectors, including industry, medical fields, auto-insurance, and image generation. In summary, LMMs have showcased their strong and diverse application capabilities in solving visual problems in the aforementioned domains. Following these existing studies, we take LMMs as the baseline model to further develop our location-integrated model, *GeoLocator*.

**2.3 Geography-related LLMs**

The use of LLMs in the spatial science domain has been relatively limited until quite recently. Earlier this year, Roberts et al. explored the geographical knowledge and reasoning skills of GPT-4 through a series of experiments, ranging from basic tasks like location estimation to complex applications like route planning and itinerary creation. Their study highlights GPT-4's capability in geospatial reasoning and its potential for diverse applications in geography-related fields (Roberts et al., 2023). Then Deng et al. developed K2, a specialized language model for geoscience, trained on a tailored corpus, showing enhanced performance in geoscience-specific tasks like question answering and knowledge reasoning, setting a new standard for domain-specific language models (Deng et al., 2023). Li et al. introduced GeoLM, a language model integrating geospatial data with linguistic information, using geo-entity anchors and spatial coordinate embeddings for enhanced geo-entities understanding (Li et al., 2023). Hu et al. developed a method that combines geo-knowledge with GPT models for improved extraction of location descriptions from social media messages during disasters. This approach, using only 22 training examples, achieved over 40% improvement in accuracy compared to standard named entity recognition methods, significantly aiding in the rapid and efficient response to disaster scenarios (Hu et al., 2023). Recently, Bhandari et al. assessed the geospatial knowledge and reasoning capabilities of LLMs, using experiments on geocoordinate prediction, geospatial preposition analysis, and multidimensional scaling, revealing their potential in geospatial reasoning tasks (Bhandari et al., 2023). Extending from the existing research, we reformulated the regular LMMs to create our location-integrated model, *GeoLocator*, and tested out its capacity to infer geospatial information and geo-privacy.

## 3. Workflow to develop the new tool – *GeoLocator*

Based on the GPT-4, we have developed a tool capable of inferring location information from images, which we have named *GeoLocator (https://chat.openai.com/g/g-qxqvMb6YJ-geolocator)*. The GPT-4 is a large multimodal model (accepting image and text inputs and emitting text outputs) that, while not as capable as a human in many real-world scenarios, has demonstrated human-level performance on a variety of professional and academic benchmarks. *GeoLocator* is a customized version of the ChatGPT that we created. *GeoLocator*'s strength is in using the powerful feature extraction and linguistic inference capabilities of large multimodal models to infer location information from images. At the same time, we developed *GeoLocator* with a large number of model commands built in to avoid transferring lengthy contexts each time. As shown in Figure 1, the most important core of *GeoLocator* is the instructions and features that can be displayed.

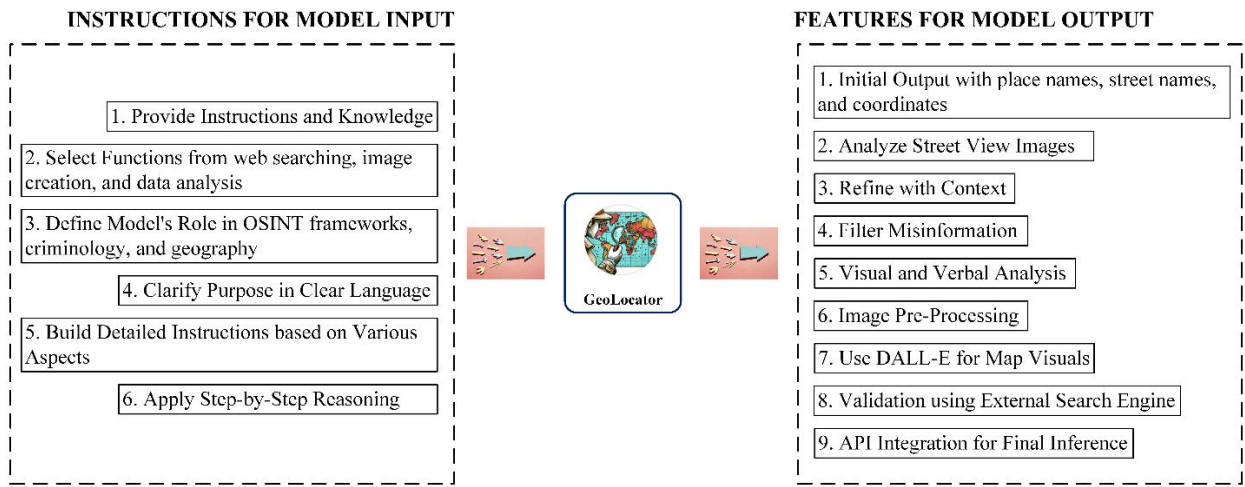

Figure 1. *GeoLocator* Instructions and Features

*GeoLocator* created based on ChatGPT (*https://chat.openai.com/gpts/editor*) with customized functionalities. Creating one is as simple as starting a conversation, giving it instructions and additional knowledge, and then choosing what it can do, such as searching the web, making images, or analyzing data. Instructions are the key to *GeoLocator*. The process of creating instructions involves engineering skills, such as defining the model's roles as being good at OSINT frameworks, criminology and geography. The purpose of the model is described in clear language and its task is to extract every detail from the photographs and come up with a sound analysis. In building the instructions for *GeoLocator*, we emphasized the need to focus on image details, EXIF data, traffic rules, human and physical geography, and unique regional clues. Step-by-step reasoning was used to improve accuracy. *GeoLocator* was built on GPT-4, which has the inherent advantage that GPT-4 can infer geographic information on its own, but the extent to which it can do so remains unknown. The *GeoLocator* we created, after enriching the directives, our *GeoLocator* has the potential to be much stronger in its ability to infer geographic information.

When built with instructions, our *GeoLocator* has a variety of features. It provides detailed place names, street names, and coordinates in the final location result. It performs an initial analysis using only street view images and refines its inferences using contextual information provided by the user. Users may provide misleading information that needs to be recognized and effectively differentiated. *GeoLocator* expresses itself visually and verbally, for example by recognizing and highlighting road signs and landmarks, and then performs a reasonably verbal analysis. It uses a code interpreter to pre-process images to deal with situations such as blurring or low resolution, including noise reduction, resolution enhancement, zooming or cropping. *GeoLocator* could utilize DALL-E image generation to obtain rich visual information, such as mapping spatial relationships (geographic manuscript style), topographic maps, or street maps. Finally, we merge the external search engine validation step into *GeoLocator* to confirm its conjectures. Using *GeoLocator'*s conclusions, we call external APIs to draw maps that provide feedback on street location information.

## 4. Experimental design to test out *GeoLocator*

We then designated a series of experiments to test out the capacity of *GeoLocator* in inferring location information and evaluating its modelling performance. We compared the results of geospatial/location information identified by three tools and/or platforms: Google search engine, GPT-4 and *GeoLocator*, based on images & texts. Such a comparison was also implemented in different languages. We observed that *GeoLocator* has the strongest location inference ability across the three tools mentioned above and can even deduce exact street addresses.

Given the size of our study, for all experiments, we judge the performance of the model solely based on whether it can correctly infer the location of the information given. To give the reader a better understanding of *GeoLocator's* analysis process for the following experiments, we have placed the flowchart shown by Figure 2.

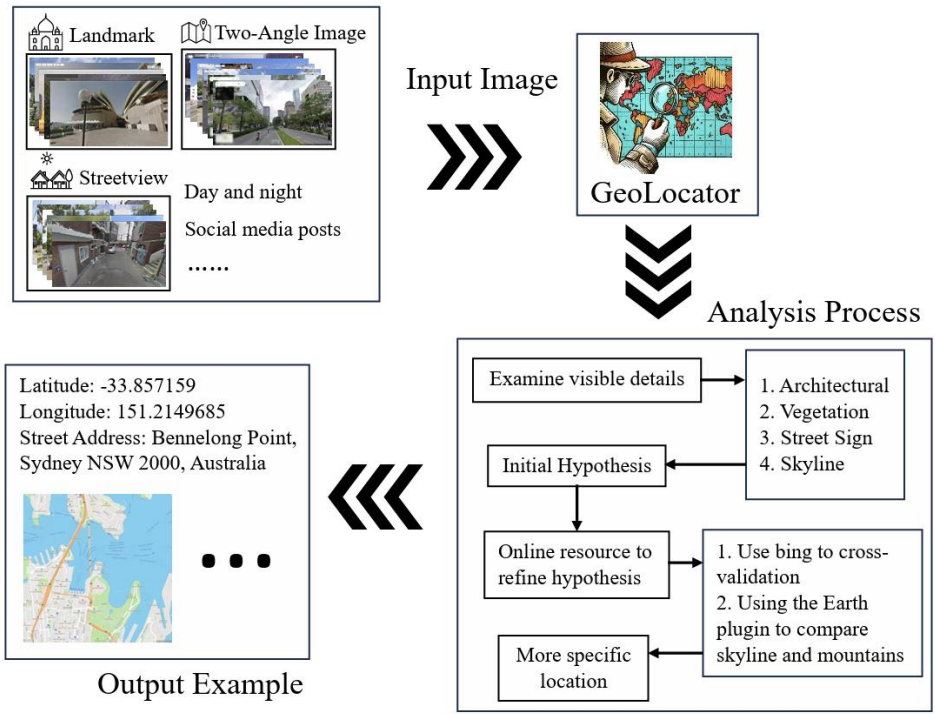

Figure 2. Geolocator working flowchart

**Step 1: Prepare the input of data**

To evaluate the effectiveness of the Google search engine, GPT-4, and *Geolocator* across various image types, we gathered a diverse set of data sources. This included images from Google Maps, photographs taken by our research team, Google Images, and posts from social media.

      Google Maps served as a resource for geographically diverse and detailed images, offering both reliability and recency. Our dataset (Table 1) comprises 100 locations from Google Maps, including 50 iconic landmarks (e.g., the Status of Liberty in New York), and 50 street views without obvious landmarks. Moreover, with the help of Google Maps, we could access different angel's street views. We selected 40 images of 20 different locations from two different angles. Additionally, the authors captured 40 images of 20 specific locations at separate times (day and night) to assess changes in environmental conditions. 10 images from Google Images were also chosen to assess the impact of language input on the results. 3 social media posts combined with text and images from the authors' personal accounts. The inclusion of social media images aimed to mimic real-life scenarios.

Table 1. Data source and description

| Data Source | Description | Number of images |
|---|---|---|
| **Google Maps** | Iconic landmarks | 50 images |
| | Street view without obvious landmarks | 50 images |
| **Taken by research team** | Images of 20 locations from two different angels | 20 sets (40 images) |
| | Images of 20 locations at two times slots (i.e., day and night time) | 20 sets (40 images) |
| **Google Images** | Images of 10 locations from China to assess the impact of language input | 10 images |
| **Posts from social media** | Social media posts sent by research team members | 3 posts |

**Step 2: Compare images' location inference ability among Google search engines, GPT-4, and GeoLocator**

In this part, we want to evaluate the location inference capabilities of Google research engines, GPT-4 and *GeoLocator* by sending photos without text prompts, thereby determining whether advanced artificial intelligence tools have better inference performance. This experiment highlights the potential of artificial intelligence tools like GPT-4 and *GeoLocator* in assisting with geographic location inference. In the experiment, we uploaded the same images, ranging from iconic landmarks to street view and day/night images of the same place, to Google search engine, GPT-4, and *GeoLocator*, and to judge the performance of the tools above based on their inference precision.

**Step 3: Compare the location inference based on image and text instruction**

Although *GeoLocator* already has high accuracy on location inference, in this experiment, we wonder whether it could perform better if we provide additional textual prompts or additional images. In this experiment, we evaluated whether *GeoLocator* could improve its performance in deducing the location of images by providing additional image perspectives or textual prompts. We selected 40 images taken from 20 different locations with different angles, and 10 images we used in step 3 with compared lower accuracy (accuracy of inference in Country level). We then compared *GeoLocator's* prediction results before and after applying these additional images or textual instructions.

**Step 4: Examine the impact of languages on inference results**

With doing more experiments, we found that *GeoLocator's* performance varies with different input languages. Therefore, we designed this experiment to evaluate the impact of different languages on the model's inference results. During the experiment, we provided the model with images containing text prompts in different languages to *GeoLocator* and observed its predictions for the geographic location of the images. And we checked if the model's predictions vary with the change of input language.

**Step 5: Evaluate the GeoLocator's performance on social media posts**

In the final experiment, we focused on a more realistic scenario, social media posts. Social media posts usually contain more complex information such as emotions, what a user is doing, and some meaningless images and text for location inferences. We explored whether finely tuned LMMs like *GeoLocator* could now infer the location of a place that social media posts show or describe.

**5. Results**

With the help of *GeoLocator*, we observed a high accuracy of the inference of location in most kinds of images mentioned in the experiment. With the advancement of technology, especially the progress in image inference and geolocation techniques, it is becoming possible to infer where a photo was taken.

**5.1 Compare images' location inference ability among Google search engines, GPT-4, and GeoLocator**

This experiment highlights the potential of artificial intelligence tools like GPT-4 and *GeoLocator*. It not only improves the speed and accuracy of inference but also expands the capabilities of users in conducting such tasks. We selected 10 representative images to show in Table 3 in detail. To simplify our results, we grouped the results of our experiments into four geographical categories (Country, State, City/Town, Street). These were color-coded for inference accuracy, ranging from dark green for the most precise to light green for the least. The outcomes of our experiments are displayed in Table 2. We consider inferences accurate to the street level as successful, and this criterion is used to evaluate the accuracy of the three tools' inferences. We calculated the image inferring accuracy by dividing the number of images successfully inferred in the experiment by the total number of images used in that kind of experiment.

      In Google search image, since the location inference of Google search engine is based on tags of images, it works well for deducing pictures of tourist attractions or iconic buildings with rich information, making it easy for search engines to infer these locations. However, it is difficult to infer the location of street views because there is little information about various street views on the internet, making it hard to establish web links and thus unable to infer street views.

      For street view image inference, GPT-4 can provide methods to predict the shooting location of images through its model used for image inference and analysis. It can analyze various factors such as landmarks, natural features, architectural styles, vegetation types, and weather conditions in the photos to complete the inference and prediction of image locations.

      Outperforming GPT-4, *GeoLocator* through our set instructions, can infer specific location information of street view images and perform a more complete inference process than GPT-4, thereby achieving better performance. It can predict not only the country or city of the

picture but even infer the name of the street or building where the image was taken without specific road signs or street names.

Table 2. Results of image inferring accuracy

| Image Type | Sample size | Google search engine | GPT-4 | GeoLocator |
|---|---|---|---|---|
| Iconic landmark | 50 | 88% | 60% | 94% |
| Street view | 50 | 16% | 18% | 54% |
| Daytime image | 20 | 25% | 40% | 70% |
| Nighttime image | 20 | 10% | 15% | 35% |

Table 3. Results of comparing images' location inference ability among Google search engines, GPT-4, and *GeoLocator*

| Images | Data Type | Google Search Engine | GPT-4 | GeoLocator | Distance* (miles) |
|---|---|---|---|---|---|
| 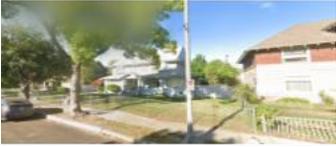 | Street View | State | City/ Town | Street | *0.0034* |
| 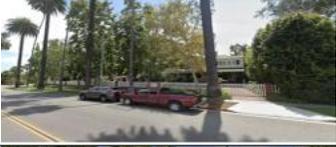 | Street View | City/ Town | Unknown** | City/ Town | *10* |
| 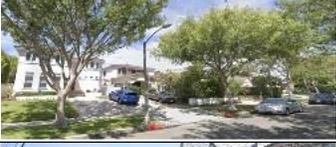 | Street View | State | Unknown** | Country | *32.49* |
| 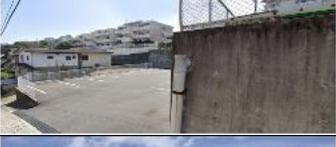 | Street View | Country | Unknown** | Country | *126.42* |
| 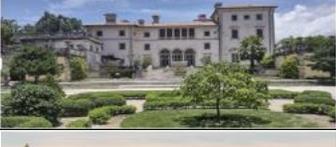 | Landmark | Street | Unknown** | Street | *0.0044* |
| 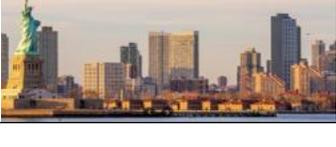 | Landmark | City/ Town | Street | Street | *0.0019* |

| 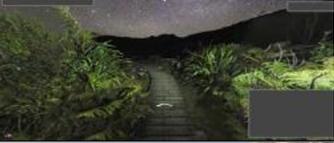 | Landmark | Unknown** | Country | City/ Town | *55.22* |
| --- | --- | --- | --- | --- | --- |
| 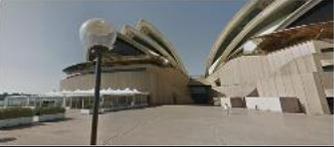 | Landmark | Street | Street | Street | *0.0449* |
| 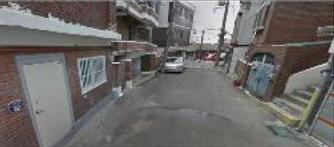 | Street View | City/ Town | Unknown** | City/ Town | *2.62* |
| 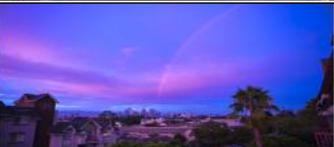 | Nighttime Image | Unknown** | Country | City/ Town | *3.28* |

Note:
*It indicates the distance between the actual address from the image and the address inferred by the GeoLocator.
**It means that the Search Engine, GPT-4, or GeoLocator is not able to infer the address.
*** Geographical Categories: Country, State, City/Town, and Street. A deeper background color indicates that it infers a more specific address in terms of the Geographical Category mentioned above.

### 5.2 Compare the location inference based on image and text instruction

We evaluated whether the *GeoLocator* could improve its performance in inferring the location of street-view images by providing additional image perspectives or textual prompts when the inference of location is not at street level. In both experiments, whether an additional image was added or there was an additional textual prompt, the model's inference accuracy increased. To assess *GeoLocator's* proficiency in inferring locations from images captured at various angles of the same site, we conducted a series of ten experiments where each scenario could potentially deduce a street-level location, using Taipei 101 as a representative example. It did not give us an expected answer at first, and we gave the model another image taken from the same place but at a different angle. Then *GeoLocator* can successfully infer the place where the photo was taken. With additional information, *GeoLocator* could get more information to infer the location further precisely in the images.

To evaluate the accuracy of location information identified by inputting additional text prompts, we test *GeoLocator's* performance. We chose to highlight the USC University Park Campus case as an example, where each scenario could potentially deduce a street-level location as well. By uploading an image representing USC University Park Campus, *GeoLocator* first observed the environment in the picture and inferred it is in a temperate of Mediterranean climate area. The people in the picture were dressed lightly and casually, and the activities looked like they were happening on a campus. Based on these clues, *GeoLocator* analyses that the picture was shot at a university in the United States. For this unsatisfactory answer, we gave the model three prompts until it gave us a street-level answer. In this situation with richer information, *GeoLocator* showed better performance that relying on image input alone. This multimodal approach enabled *GeoLocator* to infer and understand the content of images, including specific details like road signs and street names more effectively. The advantage of this multimodal analysis is that it is not just a simple overpay of different modalities of information, but rather a deep understanding and analysis of the interrelationships between these diverse types of data, leading to more comprehensive and precise conclusions. In this situation with more information, *GeoLocator* performed better than when only inputting images, being able to infer more images with specific details like road signs and street names.

### 5.3 Examine the impact of languages on inference results

In this experiment, we found different inputs will impact *GeoLocator's* reasoning procedure to affect the inference results. During the experiment, we provided the *GeoLocator* with the model with images containing text prompts in different languages and observed its reference for the geographical location of the images. For different language inputs in Chinese and English, there is no significant difference in the inference results of the model. The results showed that the model's reference indeed varies with the change of input language. For example, although the model conducted similar analyses when inputting different languages, their thought processes were different: when the input text was in English, the model prompted us with operations such as natural geography, flora, architecture, and infrastructure; whereas when the input text was in Chinese, the model prompted us with operations to refer the country, regional and zoning clues, focusing on very specific clues. Moreover, when asked in English, the model answered that the picture was from a park, which did not occur when asked in Chinese. The phenomenon may be caused by multiple factors. Firstly, the bias in the language-related datasets the model may have been exposed to during training could lead the model to associate more with images related to the geographical location of users of a particular language when processing text in that language. Secondly, texts in different languages may imply specific cultural and geographic background information, which the model might use as clues to predict the image's location.

### 5.4 Evaluate the GeoLocator's performance on social media posts

The last experiment tests the *GeoLoctor's* ability to infer the location of a picture taken in a more complex real-world scenario. Since social media posts always contain photos of the same location in different orientations, some text describing the time or place, and there contain some useless information, this requires a higher level of ability for *GeoLocator* to synthesize a location. Additionally, we asked *GeoLocator* to generate a personal profile of the person who posts, based on both the image and text. This profile includes details such as the individual's location, age, and gender. The experiment consists of 3 sub-experiments, comprising two tourist posts and one daily life post. The results show that *GeoLocator* can infer the location down to the city level and provide a detailed personal profile of the individual posting. It can precisely pinpoint the location of the exact street or area for two out of the three posts. Additionally, the distances between

*GeoLocator's* estimations and the actual locations in the images are all less than 100 miles, being 1.23, 38.01, and 0.27 miles, respectively.

**6. Discussion and conclusion**

Our development of *GeoLocator*, based on over 200 experimental tests across diverse data sources including Google Maps, author-captured images, Google Images, and social media posts, has led to four key discoveries. Firstly, *GeoLocator* demonstrated exceptional performance in inferring specific locations, particularly street views, surpassing both Google search engines and GPT-4. Secondly, we observed an enhancement in *GeoLocator*'s performance when provided with additional images from different angles or textual prompts. Thirdly, we identified that the language of input text influences *GeoLocator*'s reasoning process. Despite overall similar levels of accuracy, different languages led to variations in the model's focus and thought processes. Lastly, *GeoLocator* exhibited robust capabilities in inferring locations from social media posts, even amidst complex and diverse data such as varying orientations and accompanying social media text. These findings emphasize the advancements in AI-driven geolocation tools and their potential in various applications, from tourism to geoprivacy security detection, particularly highlighting *GeoLocator's* exceptional ability to infer locations from complex multimodal data sources.

Our development of *GeoLocator* contributes to and extends the existing research on LMMs and their applications, especially in geography-related tasks in multiple aspects. Our *GeoLocator* represents an innovative integration of geospatial data with LMMs. While existing research, like Li et al.'s work on GeoLM (Li et al., 2023), has begun exploring the integration of linguistic information with geospatial data, *GeoLocator* takes this integration further. It processes not only textual and visual data but also effectively infers detailed geospatial information such as street-level locations, enhancing the granularity of location inference. Furthermore, our *GeoLocator* extends the application of LMMs in geospatial contexts. Previous research demonstrated LMMs' capabilities in general tasks like geoscience and medical imaging. However, *GeoLocator* specifically targets the nuanced task of geoprivacy and location inference, demonstrating effectiveness in complex real-world scenarios such as social media analysis involving multiple data types and orientations. Our findings align with and diverge from existing literature, similar to studies by Roberts et al. and Bhandari et al., showing robust geospatial reasoning capabilities in LMMs (Roberts et al., 2023; Bhandari et al., 2023). However, the enhanced performance of *GeoLocator* in street-level inference and its effectiveness in multimodal analysis (including textual, visual, and geographic data) marks a significant advancement beyond the general capabilities discussed in the current literature. Our study provides new insights into the impact of language on LMMs' inference abilities. While existing studies have primarily focused on the model's performance in specific domains or tasks, our findings highlight the nuanced ways in which input language can affect a model's reasoning process and outcome, adding a new dimension to the understanding of LMMs' capabilities.

The emergence and application of our *GeoLocator* in the field of geospatial data analysis have significant policy implications, especially regarding privacy and governmental usage. We believe governments could leverage *GeoLocator* for more effective enforcement of privacy regulations. By understanding the capabilities and limitations of such advanced geospatial reasoning tools, policymakers can develop more informed guidelines and regulations to prevent and protect user geoprivacy. In urban planning and development, *GeoLocator* could assess the potential privacy impacts of new projects, such as how new constructions or urban developments might affect an area's geospatial data footprint and its implications for resident privacy. We also see great potential for *GeoLocator* in the fields of public safety and natural disaster management. In situations like natural disasters or public safety emergencies, *GeoLocator* could help governments quickly and accurately assess on-the-ground conditions, such as its potential to detect natural disaster locations like fires in real-time through social media.

We recognized certain limitations in the data and methodology used for developing and implementing *GeoLocator*. A notable limitation observed was GPT-4's handling of lengthy instructions. When provided with substantial textual input, GPT-4 often struggled with effectively processing the middle sections of the text. This raised concerns about the model's ability to consistently interpret and analyze long-form instructions or data inputs. Additionally, the variability in GPT-4's outputs underlines the need for a method to stabilize its responses, which is crucial for achieving consistent and reproducible results, especially in applications where decision-making heavily relies on the model's output. Another concern is GPT-4's permeability, particularly its tendency to inadvertently reveal its construction information and custom prompts. This poses a risk of unintentional exposure to sensitive or proprietary data. Moreover, the model's potential to provide download links for private knowledge bases is a significant security concern. Ensuring the confidentiality and integrity of data processed by GPT-4 is paramount, especially when dealing with sensitive geospatial information.

In summary, our development of *GeoLocator* represents a significant leap in the integration of language models with geospatial data, bringing to light the potential risks of geoprivacy infringement through image and text analysis. This research underscores the potential of AI-driven tools in enhancing our understanding and interaction with geographical data, while simultaneously emphasizing the need for careful consideration of privacy and ethical impacts. *GeoLocator's* ability to analyze and infer precise locations from complex multimodal data sources marks a breakthrough in the field, demonstrating the transformative power of AI in reshaping our approach to geospatial analysis. However, this innovation also carries the responsibility to ensure the ethical and secure use of such technologies. Our study is not merely a pilot exploration but also serves as a cautionary tale, setting a precedent for future research with a focus on balancing technological advancement and the preservation of geoprivacy. Imagine the unsettling realization that your exact geographical location and movements could be inferred by GPT-4 from the images and texts you post on social media, constantly under the watchful eye of AI. Therefore, as we continue to push the boundaries of what AI can achieve, remaining vigilant about the implications of these advancements is crucial, ensuring that they are used to improve society while safeguarding individual privacy and security.